\documentclass[twoside,twocolumn,9pt]{article}
\usepackage{extsizes}
\usepackage[super,sort&compress,comma]{natbib} 
\usepackage[version=3]{mhchem}
\usepackage[left=1.5cm, right=1.5cm, top=1.785cm, bottom=2.0cm]{geometry}
\usepackage{balance}
\usepackage{mathptmx}
\usepackage{sectsty}
\usepackage{graphicx} 
\usepackage{lastpage}
\usepackage[format=plain,justification=justified,singlelinecheck=false,font={stretch=1.125,small,sf},labelfont=bf,labelsep=space]{caption}
\usepackage{float}
\usepackage{fancyhdr}
\usepackage{fnpos}
\usepackage[english]{babel}
\addto{\captionsenglish}{%
  
}
\usepackage{array}
\usepackage{droidsans}
\usepackage{charter}
\usepackage[T1]{fontenc}
\usepackage[usenames,dvipsnames]{xcolor}
\usepackage{setspace}
\usepackage[compact]{titlesec}
\usepackage{hyperref}

\usepackage{epstopdf}

\definecolor{cream}{RGB}{222,217,201}

\usepackage{booktabs} 

\begin{document}

\pagestyle{fancy}
\thispagestyle{plain}
\fancypagestyle{plain}{
\renewcommand{\headrulewidth}{0pt}
}

\makeFNbottom
\makeatletter
\renewcommand\LARGE{\@setfontsize\LARGE{15pt}{17}}
\renewcommand\Large{\@setfontsize\Large{12pt}{14}}
\renewcommand\large{\@setfontsize\large{10pt}{12}}
\renewcommand\footnotesize{\@setfontsize\footnotesize{7pt}{10}}
\makeatother

\renewcommand{\thefootnote}{\fnsymbol{footnote}}
\renewcommand\footnoterule{\vspace*{1pt}%
\color{cream}\hrule width 3.5in height 0.4pt \color{black}\vspace*{5pt}} 
\setcounter{secnumdepth}{5}

\makeatletter 
\renewcommand\@biblabel[1]{#1}            
\renewcommand\@makefntext[1]%
{\noindent\makebox[0pt][r]{\@thefnmark\,}#1}
\makeatother 
\renewcommand{\figurename}{\small{Fig.}~}
\sectionfont{\sffamily\Large}
\subsectionfont{\normalsize}
\subsubsectionfont{\bf}
\setstretch{1.125} 
\setlength{\skip\footins}{0.8cm}
\setlength{\footnotesep}{0.25cm}
\setlength{\jot}{10pt}
\titlespacing*{\section}{0pt}{4pt}{4pt}
\titlespacing*{\subsection}{0pt}{15pt}{1pt}

\fancyfoot{}
\fancyfoot[LO,RE]{\vspace{-7.1pt}\includegraphics[height=9pt]{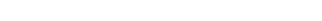}}
\fancyfoot[CO]{\vspace{-7.1pt}\hspace{13.2cm}\includegraphics{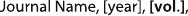}}
\fancyfoot[CE]{\vspace{-7.2pt}\hspace{-14.2cm}\includegraphics{head_foot/RF}}
\fancyfoot[RO]{\footnotesize{\sffamily{1--\pageref{LastPage} ~\textbar  \hspace{2pt}\thepage}}}
\fancyfoot[LE]{\footnotesize{\sffamily{\thepage~\textbar\hspace{3.45cm} 1--\pageref{LastPage}}}}
\fancyhead{}
\renewcommand{\headrulewidth}{0pt} 
\renewcommand{\footrulewidth}{0pt}
\setlength{\arrayrulewidth}{1pt}
\setlength{\columnsep}{6.5mm}
\setlength\bibsep{1pt}

\makeatletter 
\newlength{\figrulesep} 
\setlength{\figrulesep}{0.5\textfloatsep} 

\newcommand{\topfigrule}{\vspace*{-1pt}%
\noindent{\color{cream}\rule[-\figrulesep]{\columnwidth}{1.5pt}} }

\newcommand{\botfigrule}{\vspace*{-2pt}%
\noindent{\color{cream}\rule[\figrulesep]{\columnwidth}{1.5pt}} }

\newcommand{\dblfigrule}{\vspace*{-1pt}%
\noindent{\color{cream}\rule[-\figrulesep]{\textwidth}{1.5pt}} }

\newcommand{\silvana}[1]{\textcolor{orange}{SB: #1]}}

\makeatother

\twocolumn[
  \begin{@twocolumnfalse}
{\includegraphics[height=30pt]{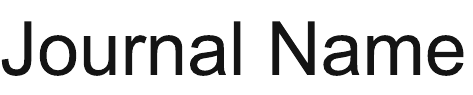}\hfill\raisebox{0pt}[0pt][0pt]{\includegraphics[height=55pt]{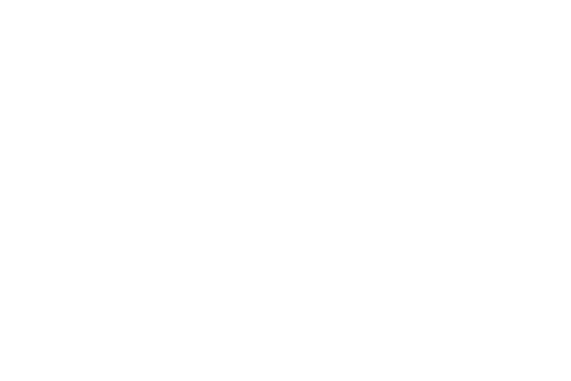}}\\[1ex]
\includegraphics[width=18.5cm]{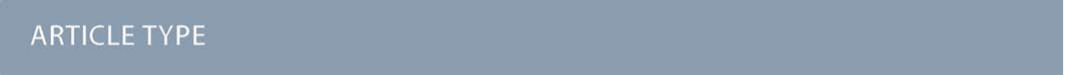}}\par
\vspace{1em}
\sffamily
\begin{tabular}{m{4.5cm} p{13.5cm} }

\includegraphics{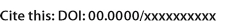} & \noindent\LARGE{\textbf{Intrinsic Point Defects and Frenkel Pair Formation in Photovoltaic Absorber Zn$_3$P$_2$: Regulating $p$-type Conductivity through Growth and Annealing Conditions$^\dag$}} \\
\vspace{0.3cm} & \vspace{0.3cm} \\

 & \noindent\large{Nico Kawashima,\textit{$^{abc}$} and Silvana Botti\,\textit{$^{ac*}$}} \\

\includegraphics{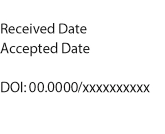} & \noindent\normalsize{%
This study investigates the ground-state energetics and thermodynamics of intrinsic point defects in zinc phosphide Zn$_3$P$_2$ using \emph{ab initio} density functional theory combined with an extensive potential energy landscape search. Our analysis reveals that the defect chemistry is dominated by zinc vacancies $V_\mathrm{Zn}$ and zinc interstitials Zn$_i$, with equilibrium concentrations significantly surpassing those of other intrinsic species. Notably, we find that phosphorus interstitials P$_i$, previously suggested to be significant, possess high formation energies and likely exist only in negligible quantities. The characteristic $p$-type conductivity of undoped Zn$_3$P$_2$ is shown to be a direct consequence of zinc vacancies, which act as shallow acceptors and pull the Fermi level toward the valence band. Furthermore, we identify a positive binding energy between $V_\mathrm{Zn}$ and Zn$_i$, leading to the formation of electrically benign Frenkel pairs that partially compensate the intrinsic p-type conductivity. Our results suggest that achieving $n$-type conductivity is fundamentally limited by these thermodynamic constraints. We conclude that hole densities can be optimized through phosphorus-rich growth conditions and high-temperature annealing, and suggest that future photovoltaic strategies should prioritize interface engineering over bulk $n$-type doping.
}

\end{tabular}

 \end{@twocolumnfalse} \vspace{0.6cm}

  ]

\renewcommand*\rmdefault{bch}\normalfont\upshape
\rmfamily
\section*{}
\vspace{-1cm}

\footnotetext{\textit{$^{a}$~Research Center Future Energy Materials and Systems of the Research Alliance Ruhr and Interdisciplinary Center for Advanced Materials Simulations, Faculty of Physics and Astronomy, Ruhr University Bochum, Universitätsstraße 150, D-44801, Bochum, Germany}}
\footnotetext{\textit{$^{b}$~ Friedrich Schiller University Jena, Institute of Condensed Matter Theory and Optics, Max-Wien-Platz 1, D-07743 Jena, Germany }}
\footnotetext{\textit{$^{c}$~European Theoretical Spectroscopy Facility (ETSF) }}
\footnotetext{* E-mail: silvana.botti@rub.de }
\footnotetext{\dag~Supplementary Information available: (...)} 

\section{Introduction}

Zinc phosphide Zn$_3$P$_2$ is a II--V semiconductor that has long been recognized as a promising candidate for thin-film photovoltaics (PV)~\cite{catalano_zn3p2_1978, pawlikowski_band_1981}. Its direct band gap of approximately $1.48$--$1.54\,\mathrm{eV}$~\cite{choi_zn3p2_2019, dimitrievska_advantage_2021, zstutz_showcasing_2022} aligns almost ideally with the solar spectrum for single-junction devices~\cite{shockley_detailed_1961}, while its high absorption coefficient allows for efficient thin-film, potentially flexible devices~\cite{swinkels_measuring_2020, micali_optimizing_2026, lemerle_nanopatterning_2026}. Beyond PV, recent studies suggest that the tunability of its electronic properties with composition may also be leveraged for thermoelectric applications~\cite{stutz_stoichiometry_2022, kawashima_computational_2026}. 

Critically, Zn$_3$P$_2$ is composed entirely of earth-abundant elements~\cite{reiss_synthesis_2016}. This makes it a highly sustainable alternative for the large-scale deployment required by rising energy demands, especially when compared to established technologies such as CdTe, Cu(In,Ga)Se$_2$ (CIGS), or GaAs, which rely on scarcer or more hazardous materials~\cite{wadia_materials_2009}. Although research into the material dates back several decades, it has seen a recent resurgence in interest driven by significant advancements in epitaxial growth, nanowire fabrication, and sophisticated thin-film deposition techniques~\cite{katsube_growth_2016, steinvall_multiple_2020, paul_van_2020, hagger_link_2026}.

Despite these promising properties, the commercial realization of Zn$_3$P$_2$-based devices remains challenging. One yet unresolved problem is the lack of a definitive consensus regarding intrinsic point defects, their impact on the electronic structure, and their influence on potential extrinsic doping strategies. While Zn$_3$P$_2$ is consistently observed to be natively $p$-type in experiments~\cite{catalano_defect_1980, weber_optical_1994, steinvall_towards_2021}, the microscopic origin of this behavior remains elusive and subject of debate. Previous theoretical and experimental investigations into its point defect chemistry have yielded inconclusive or even contradictory results~\cite{demers_intrinsic_2012, yin_electronic_2013, yuan_first-principles_2023, flor_raman_2021}. Consequently, a unified physical model is still lacking. This discrepancy is further enhanced by the wide range of off-stoichiometric stability displayed by this material~\cite{stutz_stoichiometry_2022}, which introduces substantial uncertainty when comparing pristine stoichiometric theoretical models with experimental samples that may express varying degrees of structural disorder or defect densities beyond the dilute limit.

From a computational perspective, the accuracy of Density Functional Theory (DFT) predictions for Zn$_3$P$_2$ is often limited by the large unit cell (which contains 40 atoms) and the accompanying complexity of the defect potential energy surface (PES). As highlighted by Mosquera-Lois \textit{et al.}~\cite{mosquera-lois_identifying_2023}, simple gradient-descent optimization often fails in these complex systems, as the algorithms become trapped in local minima. A common pitfall is the inadvertent enforcement of high-symmetry starting configurations. If the true ground-state configuration involves a symmetry-breaking distortion, it may remain undetected, leading to overestimated formation energies and incorrect transition levels. 

In this work, we address these challenges by conducting a rigorous and comprehensive search of the defect configurational space. Furthermore, we account for metastable configurations that may coexist in experimental samples at finite temperatures. The transition energies associated with these metastable states might directly influence carrier recombination and, ultimately, the performance limits of Zn$_3$P$_2$ solar cells~\cite{kavanagh_impact_2022, mosquera-lois_imperfections_2023}.

Finally, we extend our analysis beyond static formation energies to include the calculation of thermodynamic properties. This allows us to provide quantitative predictions of temperature-dependent and chemical potential-dependent defect concentrations. Our results both validate existing findings and reveal new, lower-energy configurations for dominant defects, resulting in significant shifts in predicted transition energies.

\section{Computational details}

The energetics of intrinsic point defects in Zn$_3$P$_2$ were calculated following established first-principles formalisms~\cite{freysoldt_first-principles_2014, goyal_computational_2017, kim_quick-start_2020}. We adhere to the modern reporting standards for computational defect studies as outlined by Squires \textit{et al.}~\cite{squires_guidelines_2026}. The formation energy $\Delta E^q_{f}$ of a defect in charge state $q$ is defined as
\begin{equation} \label{equ:defect_formation_energy}
    \Delta E^q_f = E_q - E_\mathrm{bulk} - \sum \limits_i n_i \mu_i + q(E_\mathrm{VBM} + \epsilon_F) + E_\mathrm{corr}\,,
\end{equation}
where $E_q$ and $E_\mathrm{bulk}$ are the total energies of the supercell containing the defect and the pristine bulk reference, respectively. The variable $n_i$ represents the number of atoms of species $i$ added ($n_i > 0$) or removed ($n_i < 0$) to create the defect, with $\mu_i$ being the corresponding chemical potential. The Fermi level $\epsilon_F$ is referenced to the valence band maximum $E_\mathrm{VBM}$ of the bulk, and $E_\mathrm{corr}$ accounts for finite-size corrections.

To approximate the dilute limit while maintaining computational feasibility, we employed an 80-atom supercell (twice the volume of the 40-atom tetragonal unit cell). This configuration ensures a minimum periodic image separation of approximately 11.3\,\AA, which we validated as sufficient to accommodate local structural distortions.

To rigorously identify the global ground-state configurations, we utilized the \texttt{ShakeNBreak} and \texttt{doped} software packages~\cite{mosquera-lois_shakenbreak_2022, kavanagh_doped_2024}. This involved a systematic search of the configuration space, which samples the potential energy surface through chemically-guided distortions and explicitly breaks local symmetry via random atomic ``rattling''. Such a treatment was shown to be necessary for overcoming the limitations of standard local optimization, which often fails to capture significant reconstructive relaxations in semiconductor defects~\cite{mosquera-lois_identifying_2023}.

We considered a comprehensive set of intrinsic point defects, including vacancies ($V_\mathrm{Zn}$, $V_\mathrm{P}$), antisites (Zn$_\mathrm{P}$, P$_\mathrm{Zn}$), and interstitials (Zn$_i$, P$_i$). For each defect type, we investigated a wide range of charge states to capture all potentially relevant transition levels within the band gap. Zinc vacancies $V_\mathrm{Zn}$ were evaluated in charge states ranging from $+1$ to $-2$, while phosphorus vacancies $V_\mathrm{P}$ spanned $+3$ to $-1$. Antisite defects Zn$_\mathrm{P}$ and P$_\mathrm{Zn}$ were investigated in ranges $[0, +5]$ and $[-5, +3]$, respectively. Zinc interstitials Zn$_i$ were treated in states $[0, +2]$, whereas the complex potential energy surface of phosphorus interstitials P$_i$ was explored across a wide range of nine charge states from $-3$ to $+5$.

The configuration space for defect complexes is inherently high-dimensional. In this work, we focused on the interaction between the most prevalent species: the zinc vacancy $V_\mathrm{Zn}$ and the zinc interstitial Zn$_i$. Rather than an exhaustive global search, we constructed a Frenkel pair by placing the herein identified lowest-energy $V_\mathrm{Zn}$ and Zn$_i$ configurations within the same supercell at a distance of about 2.6\,\AA, followed by a full geometry optimization to determine the stable local geometry and binding energy. Charge states ranging from $-2$ to $+2$ were considered.

To model experimentally relevant growth environments, chemical potential limits were derived from the calculated Zn--P binary phase diagram. Relevant competing materials and phases were derived from the Materials Project~\cite{jain_commentary_2013} alongside results from our earlier work~\cite{kawashima_computational_2026}. We considered both Zn-rich and P-rich chemical potential extremes, where the accessible range is defined by equilibrium with elemental Zn (Zn-rich limit) and ZnP$_2$ (P-rich limit).

Spurious electrostatic interactions between charged periodic images were addressed using the Kumagai-Oba correction scheme~\cite{kumagai_electrostatics-based_2014}. This approach, extending the Freysoldt-Neugebauer-Van de Walle formalism~\cite{freysoldt_fully_2009}, utilizes the anisotropic static dielectric tensor to model the defect-induced potential.

Total energy calculations, structural relaxations, and the computation of dielectric properties were performed using Density Functional Theory (DFT) as implemented in the \texttt{VASP} code~\cite{kresse_efficient_1996, kresse_efficiency_1996} (version 6.3.2) within the Projector Augmented Wave (PAW) formalism~\cite{blochl_projector_1994}. The valence electron configurations were treated as $d^{10}p^2$ for Zn (12 electrons) and $s^2p^3$ for P (5 electrons).

Initial structural relaxations of defect supercells were performed using the PBEsol functional~\cite{perdew_generalized_1996, perdew_restoring_2008}, while final electronic properties and total energies were obtained using the HSE06 screened hybrid functional~\cite{heyd_hybrid_2003} with the standard mixing parameter ($\alpha = 0.25$). All calculations were spin-polarized (collinear) and included non-spherical contributions to the gradient of the density within the PAW spheres. While the often cited experimental band gap of $E_\mathrm{gap} = 1.51\,\mathrm{eV}$ was utilized for post-calculation alignment of transition levels (scissoring), the internal electronic structure was calculated using the default HSE06 parameters.

The structural model for tetragonal Zn$_3$P$_2$ was sourced from the Materials Project database (\texttt{mp-2071})~\cite{jain_commentary_2013}. Bulk lattice parameters were subsequently optimized using the PBEsol functional, as detailed in Section 1 of the Supplementary Information (SI). For all subsequent defect calculations, the supercell volume and shape were fixed at these bulk values while the internal ionic positions were relaxed. Although the final supercell energies were computed using the hybrid HSE06 functional on PBEsol-optimized geometries, the resulting errors from spurious strain and residual internal forces~\cite{gouveia_can_2019} are considered negligible.

A $\Gamma$-centered $3 \times 3 \times 3$ $k$-point mesh was used to sample the Brillouin zone of supercells. For the plane-wave basis we set an energy cutoff of $400\,\mathrm{eV}$. Electronic cycles were converged to $2\times 10^{-5}\,\mathrm{eV}$, and ionic relaxations were continued until all interatomic forces were below $10\,\mathrm{meV}/$\AA. 

A precise evaluation of the pristine density of states (DOS) is needed for determining charge carrier concentrations, Fermi level, and equilibrium defect concentrations. To this end, DOS calculations were performed using the HSE06 functional on the PBEsol-optimized primitive unit cell. To ensure numerical accuracy, a dense $k$-point mesh was employed in conjunction with tetrahedron smearing. All remaining parameters were kept consistent with defect supercell calculations.

The thermodynamic stability of the Zn--P system was assessed by considering the competing on-hull phases Zn, P, and ZnP$_2$. Following a computational workflow consistent with the aforementioned defect studies, all phases were relaxed at the PBEsol level before final energy evaluation with the HSE06 functional.

The macroscopic static dielectric tensor of Zn$_3$P$_2$ was determined by accounting for both electronic and ionic contributions. The high-frequency (optical) dielectric constant was calculated in the independent particle approximation using the HSE06 hybrid functional, incorporating spin-orbit coupling (SOC) effects. These calculations employed a plane-wave energy cutoff of $350\,\mathrm{eV}$ and a $\Gamma$-centered $5\times 5\times 5$ $k$-point grid. To ensure convergence of the dielectric response within the density-density framework, 940 bands were included in the summation.

The ionic contribution to the static dielectric constant was determined using the PBEsol functional, also with the inclusion of SOC. For these calculations, a higher plane-wave energy cutoff of $700\,\mathrm{eV}$ and a $\Gamma$-centered $5\times 5\times 4$ $k$-point grid were utilized to accurately capture the macroscopic static dielectric tensor. The final total dielectric constant used for subsequent defect formation energy corrections was derived from the combination of these electronic and lattice components.

The equilibrium concentration $[D^q]$ of a defect in charge state $q$ at temperature $T$ is determined by its formation energy $\Delta E^q_{f}$ through the Boltzmann distribution
\begin{equation}\label{equ:defect_conc_from_E}
    [D^q] = N_s g_q \exp \left( -\frac{\Delta E^q_{f}(\epsilon_F)}{k_B T} \right)\,,
\end{equation}
where $N_s$ is the density of possible site positions in the lattice, $g_q$ is the configurational and spin degeneracy, and $k_B$ is the Boltzmann constant. To determine the equilibrium Fermi level $\epsilon_F$ and the self-consistent carrier concentrations, we enforce the macroscopic charge neutrality condition,
\begin{equation}\label{equ:charge_neutrality_cond}
    n_0(\epsilon_F) - p_0(\epsilon_F) + \sum_{i, q} q [D_i^q] = 0\,,
\end{equation}
where $n_0$ and $p_0$ are the free electron and hole concentrations, respectively. This system of equations is solved iteratively, as the concentrations of charged defects depend implicitly on $\epsilon_F$, which is in turn constrained by the total charge density.

For defect complexes, we further calculated the binding energy $E_b$,
\begin{equation}\label{equ:binding_energy_complex}
    E_b = \Delta E^f[A] + \Delta E^f[B] - \Delta E^f[AB]\,,
\end{equation}
where $\Delta E^f[A]$ and $\Delta E^f[B]$ are the formation energies of the isolated constituents, and $\Delta E^f[AB]$ is the formation energy of the complex. A positive value for $E_b$ denotes an attractive interaction and the potential for stable complex formation.

\section{Results and Discussion}

\subsection{Pristine zinc phosphide}

We start by describing the properties of the pristine crystal without any defects, as determined by this work.

\begin{figure}
\centering 
      \includegraphics[height=6cm]{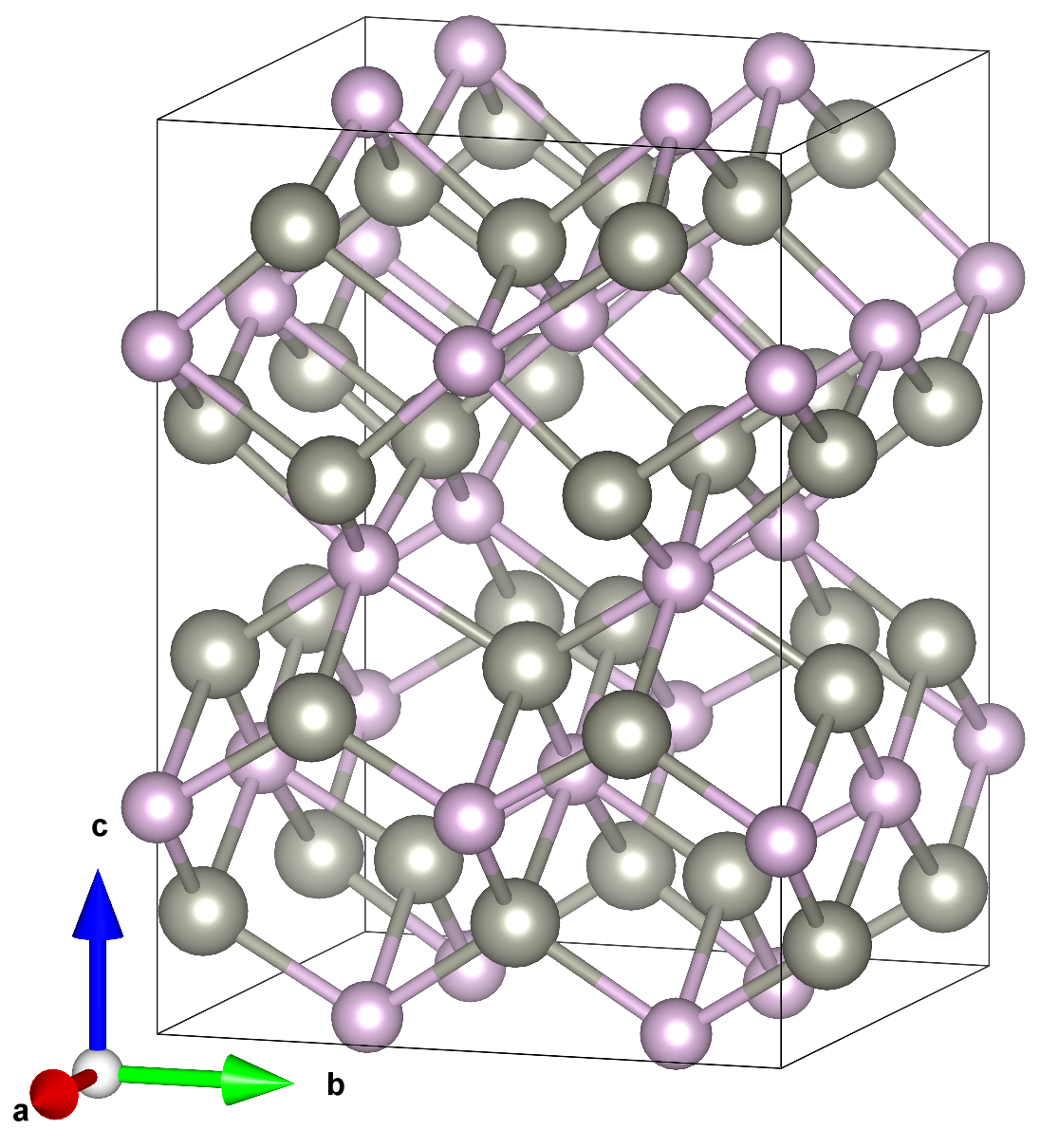} 
      \caption{\ Pristine tetragonal unit cell of zinc phosphide Zn$_3$P$_2$ containing 24 Zn-atoms (gray) and 16 P-atoms (purple).}
      \label{fig:Zn3P2_pristine_uc}
\end{figure}

In its ground state, zinc phosphide has a complex structure expressed by its large tetragonal unit cell which contains 24 zinc and 16 phosphorous atoms, see Fig.~\ref{fig:Zn3P2_pristine_uc}. The structure is invariant under the 16 symmetry operations of the group P4$_2$/nmc (\#137), which includes point inversion at the center of the unit cell.

The choice of functional for \emph{ab initio} geometry optimization significantly influences the predicted unit cell dimensions. In Section 1 of the SI, we provide a comparison of unit cell parameters obtained via gradient descent calculations using three distinct functionals: LDA, PBE, and PBEsol. We chose to proceed with the PBEsol-optimized model for all further point-defect calculations.

\begin{figure}
    \centering
    \includegraphics[width=1.0\linewidth]{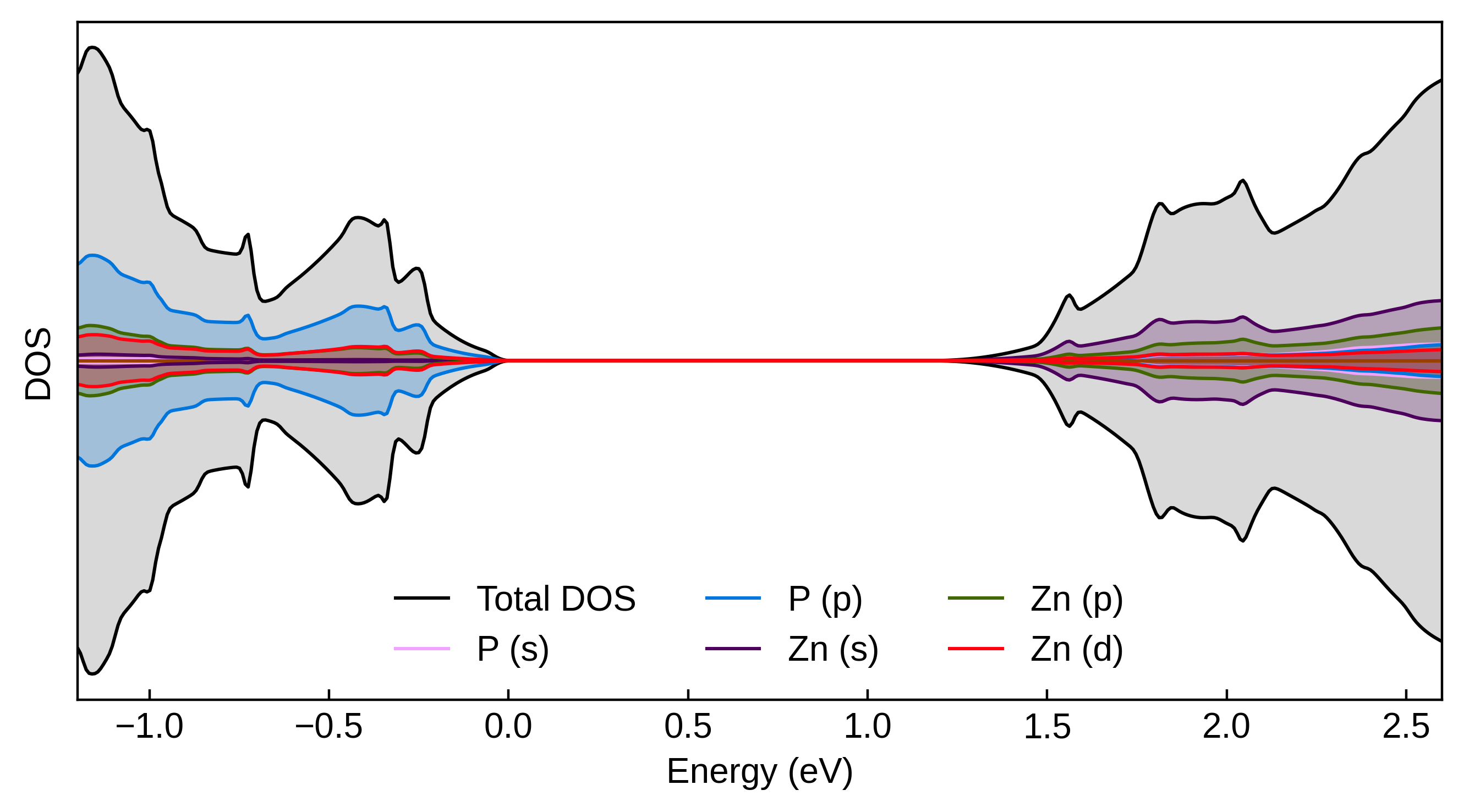} 
    \caption{\ Close-up of the total and projected DOS of Zn$_3$P$_2$ near the band gap, computed using the HSE06 functional. The spin-resolved DOS is shown, with spin-up states plotted with positive intensity and spin-down states plotted with negative intensity.}
    \label{fig:Zn3P2_DOS}
\end{figure}

\begin{figure}
    \centering
    \includegraphics[width=0.9\linewidth]{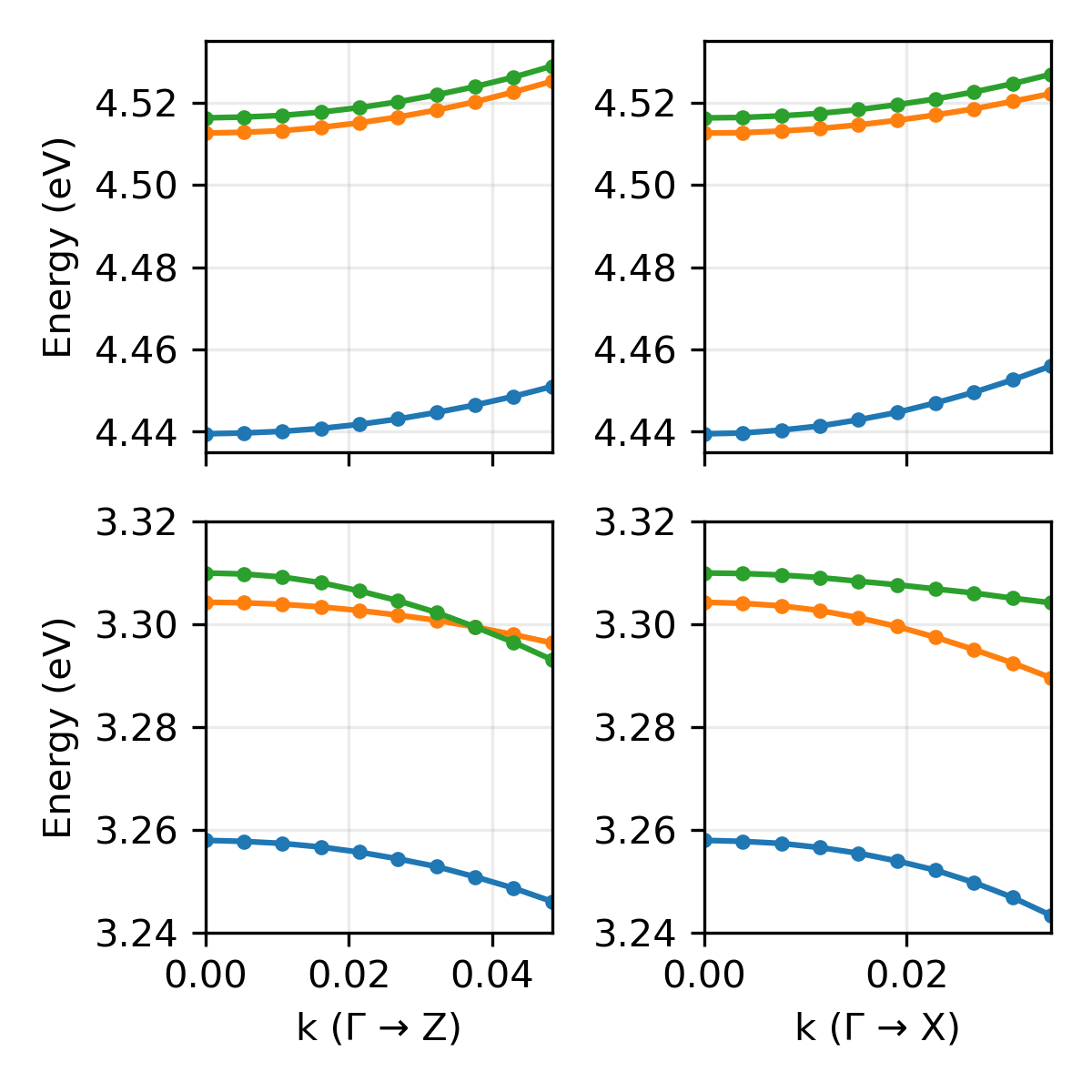}
    \caption{\ Electronic band structure of Zn$_3$P$_2$ near the band edges. Valence bands (bottom) and conduction bands (top) are shown along the $\Gamma$-Z (left) and $\Gamma$-X (right) directions. Effective masses reported in Table~\ref{tbl:Zn3P2_pristine_eff_masses} were obtained by fitting the band curvature in the vicinity of $\Gamma$.}
    \label{fig:Zn3P2_pristine_bands_closeup_effective_mass_fits}
\end{figure}

The band gap as computed via HSE06 was determined to be $1.16\,\mathrm{eV}$, which is below the experimentally determined $1.48$--$1.54\,\mathrm{eV}$~\cite{choi_zn3p2_2019, dimitrievska_advantage_2021, zstutz_showcasing_2022}. An underestimation of this magnitude is consistent with the documented performance of HSE06 for semiconductors~\cite{borlido_large-scale_2019, borlido_exchange-correlation_2020}. In Fig.~\ref{fig:Zn3P2_DOS} we show the density of states near the band gap as calculated with the HSE06 functional. We found little impact of spin-orbit coupling on the magnitude of the band gap.
We also took a close look at effective masses and determined their values from fitting 3rd-order polynomials. See Fig.~\ref{fig:Zn3P2_pristine_bands_closeup_effective_mass_fits} for a closeup of the fitted band structure and Table~\ref{tbl:Zn3P2_pristine_eff_masses} for the final results. We note that all these values, but especially the heavy hole, come with some uncertainty. A discussion of the reported errors can be found in Section 2 of the SI. The influence of the choice of functional is discussed in Section 3 of the SI.
\begin{table}
\small
    \caption{\ Effective carrier masses at the $\Gamma$ point, given in units of electron mass $m_0$. Bands are classified as valence (VB) or conduction (CB) bands and indexed by their relative energies at $\Gamma$; the index 1 denotes the valence band maximum or conduction-band minimum, respectively. By convention, negative effective masses correspond to valence band (hole) states.}
    \label{tbl:Zn3P2_pristine_eff_masses}
    \begin{tabular*}{0.48\textwidth}{@{\extracolsep{\fill}}llr}
        \toprule
        path    & band  & effective mass ($m_0$) \\
        \midrule
        $\Gamma\ \rightarrow$ X & VB3   & -0.26 $\pm$ 0.09  \\  
                                & VB2   & -0.24 $\pm$ 0.10  \\  
                                & VB1   & -0.50 $\pm$ 0.20  \\  
                                & CB1   &  0.17 $\pm$ 0.01  \\  
                                & CB2   &  0.28 $\pm$ 0.02  \\  
                                & CB3   &  0.27 $\pm$ 0.02  \\  
        $\Gamma\ \rightarrow$ Z & VB3   & -0.31 $\pm$ 0.09  \\  
                                & VB2   & -0.35 $\pm$ 0.01  \\  
                                & VB1   & -0.16 $\pm$ 0.02  \\  
                                & CB1   &  0.31 $\pm$ 0.09  \\  
                                & CB2   &  0.27 $\pm$ 0.05  \\  
                                & CB3   &  0.27 $\pm$ 0.05      
    \end{tabular*}
\end{table}

Finally, we computed the dielectric tensor of pristine Zn$_3$P$_2$. This is an important additional characterization as the material is proposed as a photovoltaic absorber. Moreover, the knowledge of the static dielectric tensor is needed for the charge corrections when computing charged defects using the supercell approach. We find, and utilize, the values for the static dielectric tensor as reported in eqs.~\eqref{equ:epsilon_opt_and_ion} and \eqref{equ:epsilon_total}:
\begin{equation} \label{equ:epsilon_opt_and_ion}
    \epsilon_\mathrm{optic} = 
    \begin{pmatrix}
        9.8 & 0     & 0     \\
        0   & 9.8   & 0     \\
        0   & 0     & 9.9   
    \end{pmatrix}\,,\ \ \ 
    \epsilon_\mathrm{ionic} = 
    \begin{pmatrix}
        54.2    & 0     & 0     \\
        0       & 54.2  & 0     \\
        0       & 0     & 12.2   
    \end{pmatrix}\,,
\end{equation}
\begin{equation} \label{equ:epsilon_total}
    \epsilon_\mathrm{total} = 
    \begin{pmatrix}
        64.0    & 0     & 0     \\
        0       & 64.0  & 0     \\
        0       & 0     & 22.1   
    \end{pmatrix}
\end{equation}

\subsection{Intrinsic point defects}

We now assess the intrinsic point defects in Zn$_3$P$_2$, including vacancies, interstitials, and antisites. Formation energies and charge-state transition levels were computed as a function of the Fermi level according to eq.~\eqref{equ:defect_formation_energy}.

The atomic chemical potentials used to calculate defect formation energies are summarized in Table~\ref{tbl:chempots}. These values represent the thermodynamic stability limits of Zn$_3$P$_2$ in equilibrium with either Zn (Zn-rich) or ZnP$_2$ (P-rich). All potentials are reported relative to the total energies of the elemental bulk phases, which were determined to be $-1.256\,\mathrm{eV/atom}$ for Zn and $-6.495\,\mathrm{eV/atom}$ for P.
\begin{table}
\centering
\caption{\small\ Summary of atomic chemical potentials ($\mu_i$) at the phase boundaries of Zn$_3$P$_2$. The Zn-rich and P-rich limits are defined by equilibrium with elemental Zn and the ZnP$_2$ secondary phase, respectively. All values are reported in $\mathrm{eV/atom}$.}
\label{tbl:chempots}
\begin{tabular}{lrr}
 & Zn--Zn$_3$P$_2$ & Zn$_3$P$_2$--ZnP$_2$ \\
\midrule
Zn & 0.000 & -0.194 \\
P & -0.638 & -0.346
\end{tabular}
\end{table}

\begin{figure}
\centering
    \includegraphics[height=5.9cm, trim={0.2cm 0 0.1cm 0}, clip]{
    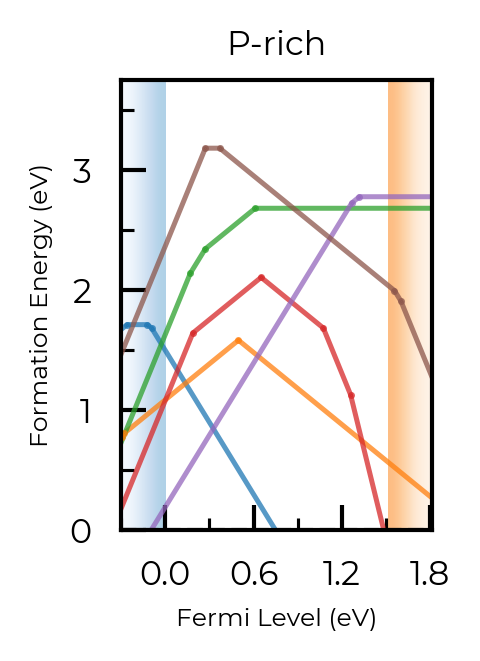}
    \includegraphics[height=5.9cm, trim={0.1cm 0 0.2cm 0}, clip]{
    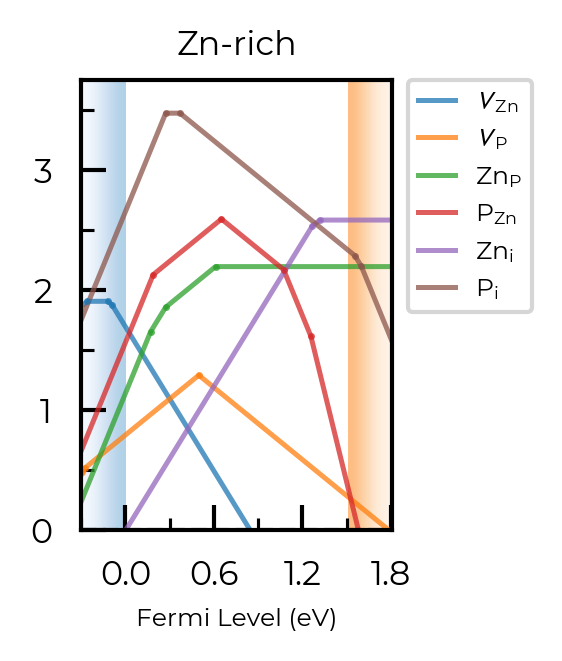}
    \caption{\ Transition energy diagrams from calculated formation energies as a function of Fermi level. Only the minimum energy configurations for each species are displayed, representing the thermodynamic ground state for each charge state.}
    \label{fig:intrinsic_pd_transition_energy_diag}
\end{figure}

The central results of this study are the transition-energy diagrams for P-rich/Zn-poor and P-poor/Zn-rich growth conditions shown in Fig.~\ref{fig:intrinsic_pd_transition_energy_diag}. For each defect type, these plots show the lowest formation energy as a function of the Fermi level across all charge states. Kinks in the curves mark thermodynamic charge-state transition levels at which the defect changes its stable charge state. A full enumeration of various computed formation energies is given in Section 4 of the SI.

We find that the Fermi level is restricted to an interval extending from the valence band maximum (VBM) to approximately $0.75$--$0.85\,\mathrm{eV}$ above it, depending on the chemical potentials of the atomic species. Outside this range, either doubly negatively charged zinc vacancies or doubly positively charged zinc interstitials acquire negative formation energies and thus form spontaneously in large concentrations, compensating free carriers and driving the Fermi level back into this window. This intrinsic restriction for the Fermi level directly implies macroscopic $p$-type behavior.

Within this Fermi level range, zinc vacancies and zinc interstitials exhibit the lowest formation energies and therefore dominate the intrinsic defect population. Phosphorus vacancies and antisite defects have higher formation energies and are expected in minor concentrations, while phosphorus interstitials are energetically least favorable and thus essentially negligible.

The influence of the chemical potentials is comparatively modest due to the limited stability range imposed by competing phases. Nevertheless, under P-rich (Zn-poor) conditions, zinc vacancies are further stabilized while zinc interstitials become less favorable. The opposite trend occurs under Zn-rich conditions. Within the Fermi level range, the relative stability of antisite defects, however, depends strongly on growth conditions. Zn$_\mathrm{P}$ antisites are favored under Zn-rich conditions, whereas P$_\mathrm{Zn}$ antisites are more favorable under P-rich conditions.

\begin{figure}
    \centering
    \includegraphics[width=1.0\linewidth]{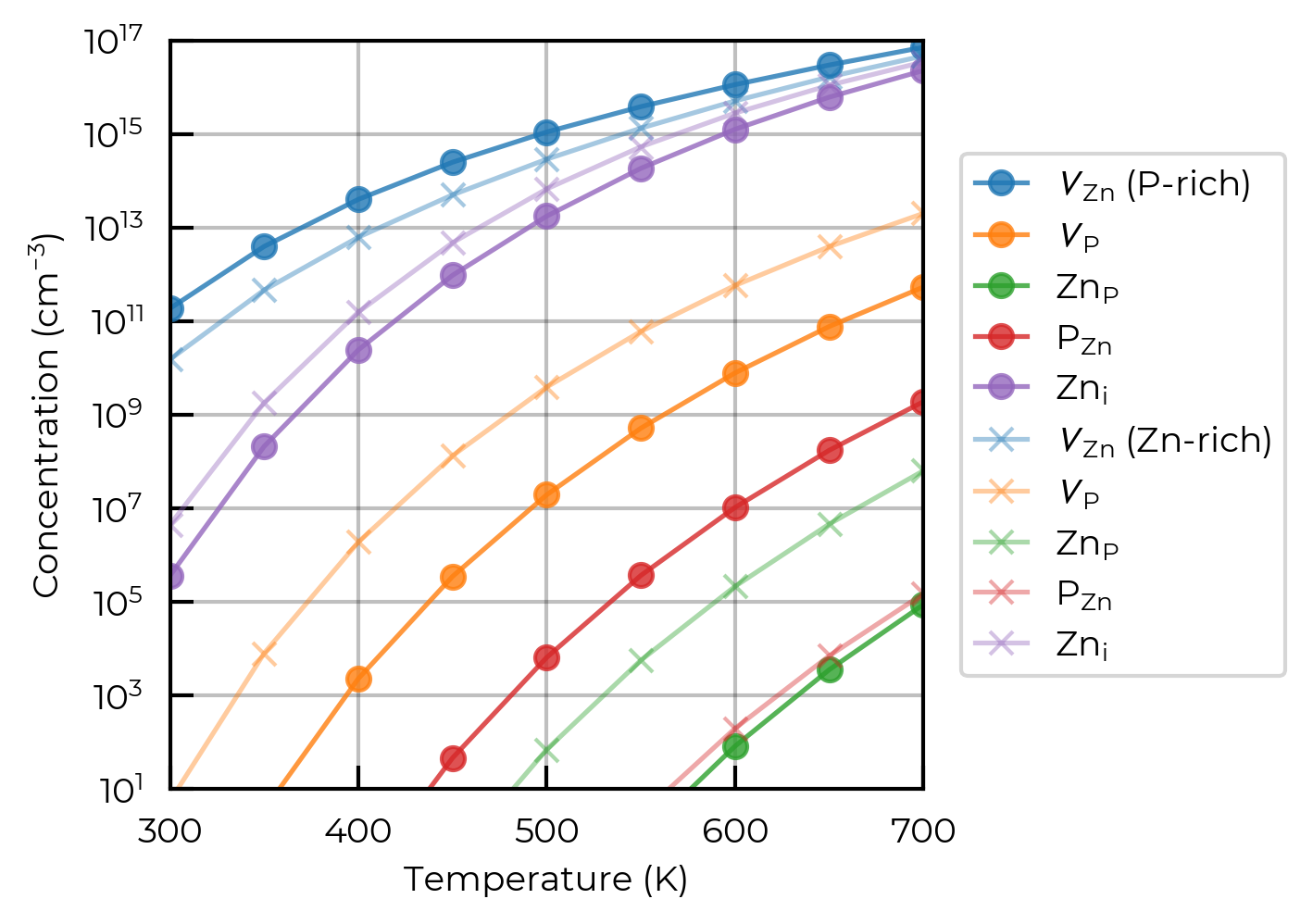}
    \caption{\ Defect concentrations at given temperature, calculated by using the computed formation energies shown in Fig.~\ref{fig:intrinsic_pd_transition_energy_diag}, assuming a Boltzmann distribution, and enforcing charge neutrality.\label{fig:defect_conc_over_T}}
\end{figure}

Using a Boltzmann formalism, i.e.\ by relying on eqs.~\eqref{equ:defect_conc_from_E} and \eqref{equ:charge_neutrality_cond}, we compute equilibrium defect concentrations and the corresponding Fermi level self-consistently as a function of temperature. Here, defect multiplicities are directly included and give a much more quantitative insight than the previous discussion based purely on the transition energy diagram. The results for Zn-rich and P-rich (Zn-poor) conditions are shown in Fig.~\ref{fig:defect_conc_over_T}. As discussed, zinc vacancies and zinc interstitials dominate overwhelmingly, with concentrations several orders of magnitude higher than all other defects, reaching values up to $10^{17}\,\mathrm{cm}^{-3}$. In contrast, phosphorus interstitials remain negligible ($<10^{1}\,\mathrm{cm}^{-3}$) due to their high formation energies and are therefore omitted from the figure. This finding contrasts with earlier studies that proposed phosphorus interstitials as dominant defects~\cite{demers_intrinsic_2012,stutz2022stoichiometry,paul2023Zn}. We return to this comparison in a later section.

\begin{figure}
    \centering
    \includegraphics[width=1.0\linewidth]{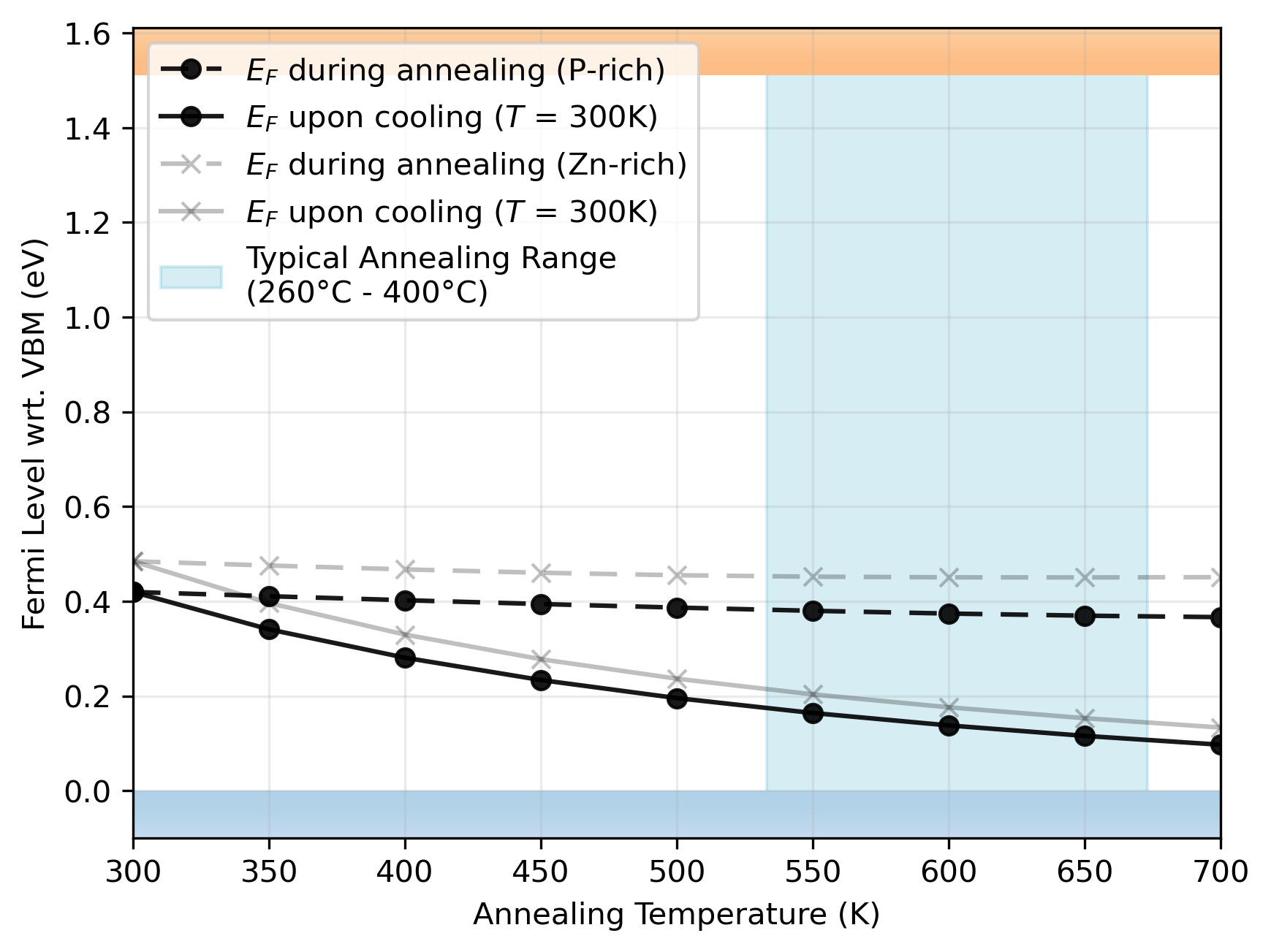}
    \caption{\ Effects of annealing at different temperatures on the equilibrium Fermi level, for samples grown under P- and Zn-rich conditions.\label{fig:E_fermi_at_annealTemp}}
\end{figure}

The self-consistent calculation of defect concentrations also yields the equilibrium Fermi level, shown in Fig.~\ref{fig:E_fermi_at_annealTemp} (dashed lines). Under both Zn-rich and Zn-poor conditions, the formation of negatively charged zinc vacancies pins the Fermi level close to the valence band maximum by binding free electrons.

In real samples, however, defect concentrations are rarely according to thermodynamic equilibrium at their operating temperature. Instead, defect populations are typically set during growth or annealing and subsequently frozen in upon cooling. They can, however, adopt equilibrium charge state distributions, as these are associated with no or only minor changes of the lattice. Within this frozen-defect approximation, only charge states re-equilibrate at the operating temperature, and one can compute the defect concentrations more realistically based on their growth or annealing history. The resulting Fermi level as a function of annealing/growth temperature is shown in Fig.~\ref{fig:E_fermi_at_annealTemp} (solid lines) for Zn-rich and Zn-poor growth conditions. As can be seen, the Fermi level shifts even closer to the valence band with increasing annealing temperature, further reinforcing the intrinsic $p$-type character. For typical growth parameters (P-rich, $T_\mathrm{anneal}=650\,\mathrm{K}$, $T_\mathrm{quench}=300\,\mathrm{K}$), we find an equilibrium Fermi level of $0.12\,\mathrm{eV}$ above the VBM.

\begin{figure}
    \centering
    \includegraphics[width=1.0\linewidth]{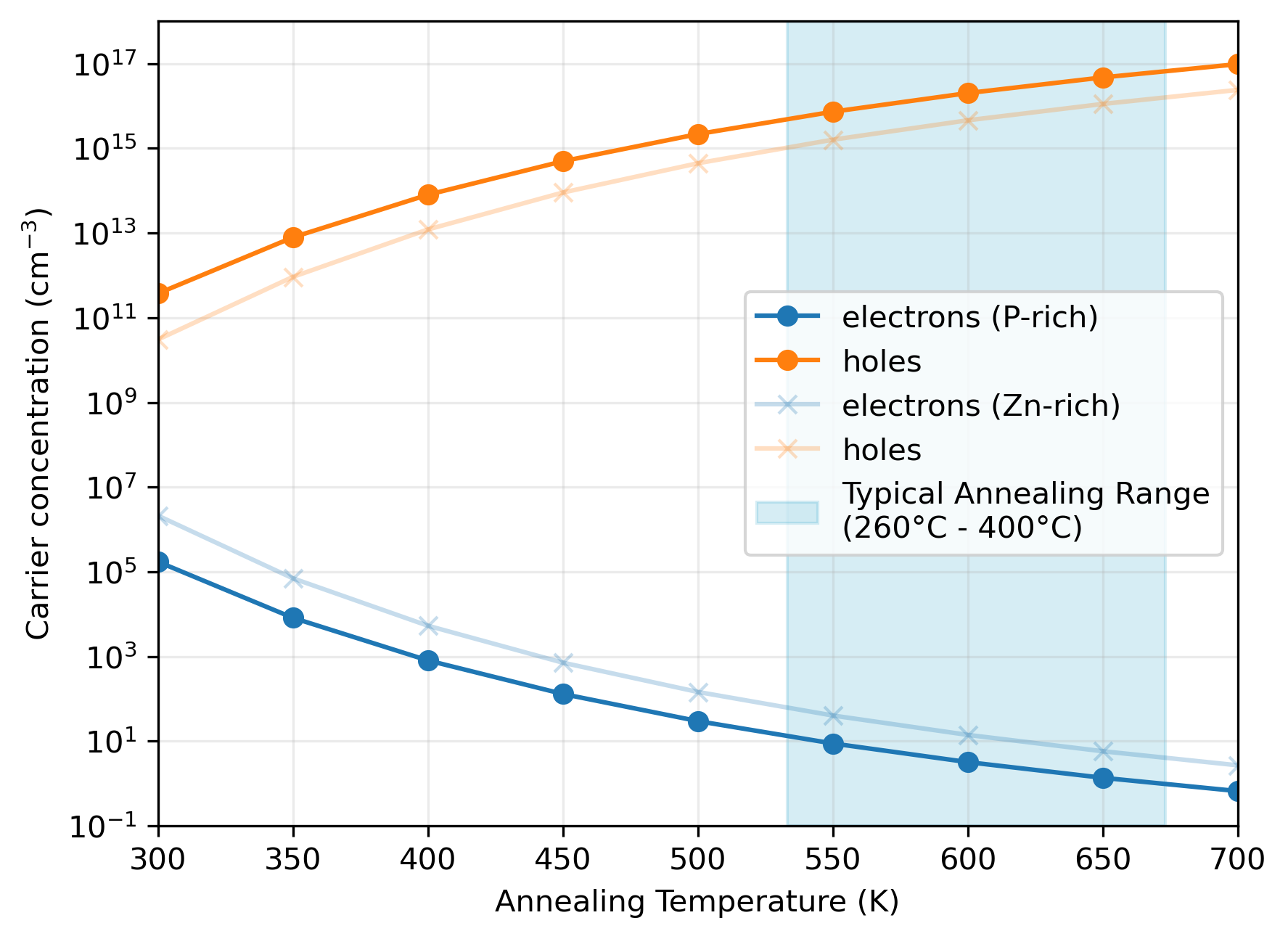}
    \caption{\ Effect of annealing at different temperatures on the final carrier concentration \emph{after} quenching to $T=300\,\mathrm{K}$, for samples grown under P- and Zn-rich conditions.\label{fig:carrier_conc_at_annealTemp}}
\end{figure}

From the knowledge of Fermi level, the density of states (especially near the band gap) and relying on the charge neutrality condition, we can explicitly compute the carrier concentration for both holes and electrons. The resulting trends are shown in Fig.~\ref{fig:carrier_conc_at_annealTemp}. As the Fermi level approaches the valence band with increasing annealing temperature, the hole concentration increases by several orders of magnitude, reaching up to $10^{17}\,\mathrm{cm}^{-3}$, in good agreement with experimental reports. This does not only give validation to our assumptions made, but it also gives a direct recipe for maximizing hole concentration for potential device design: To increase the hole concentration based on point defect thermodynamics, our results suggest that P-rich growth conditions and high growth temperature, or alternatively post-growth annealing at high temperatures, will lead to the highest hole concentrations.

Our results also clarify the long-standing difficulty of achieving $n$-type Zn$_3$P$_2$~\cite{kimball_mg_2010}. Raising the Fermi level toward the conduction band increases the formation of compensating acceptor defects, primarily negatively charged zinc vacancies. In addition, charge redistribution during cooling further traps electrons in negatively charged defects. The most favorable conditions for electron carriers are Zn-rich growth and low-temperature processing or slow cooling, which maintain defect populations closer to equilibrium at operating conditions. Even under these conditions, however, the predicted concentrations remain too low for electrons to become the majority carriers.

Transition-energy diagrams also provide information on defect levels within the band gap. Here, one should distinguish thermodynamic transition levels, obtained from formation-energy differences, from optical transition levels. The energy levels within the band gap which can be directly read off the transition energy diagrams like the ones in shown in Fig.~\ref{fig:intrinsic_pd_transition_energy_diag} correspond to those thermodynamic transitions that would become apparent in equilibrium measurements such as deep-level transient spectroscopy. Optical transitions (e.g., PL or CL) occur on timescales too short for full lattice relaxation and are therefore shifted relative to thermodynamic levels by the Stokes shift.

\begin{figure}
    \centering
    \includegraphics[width=1.0\linewidth]{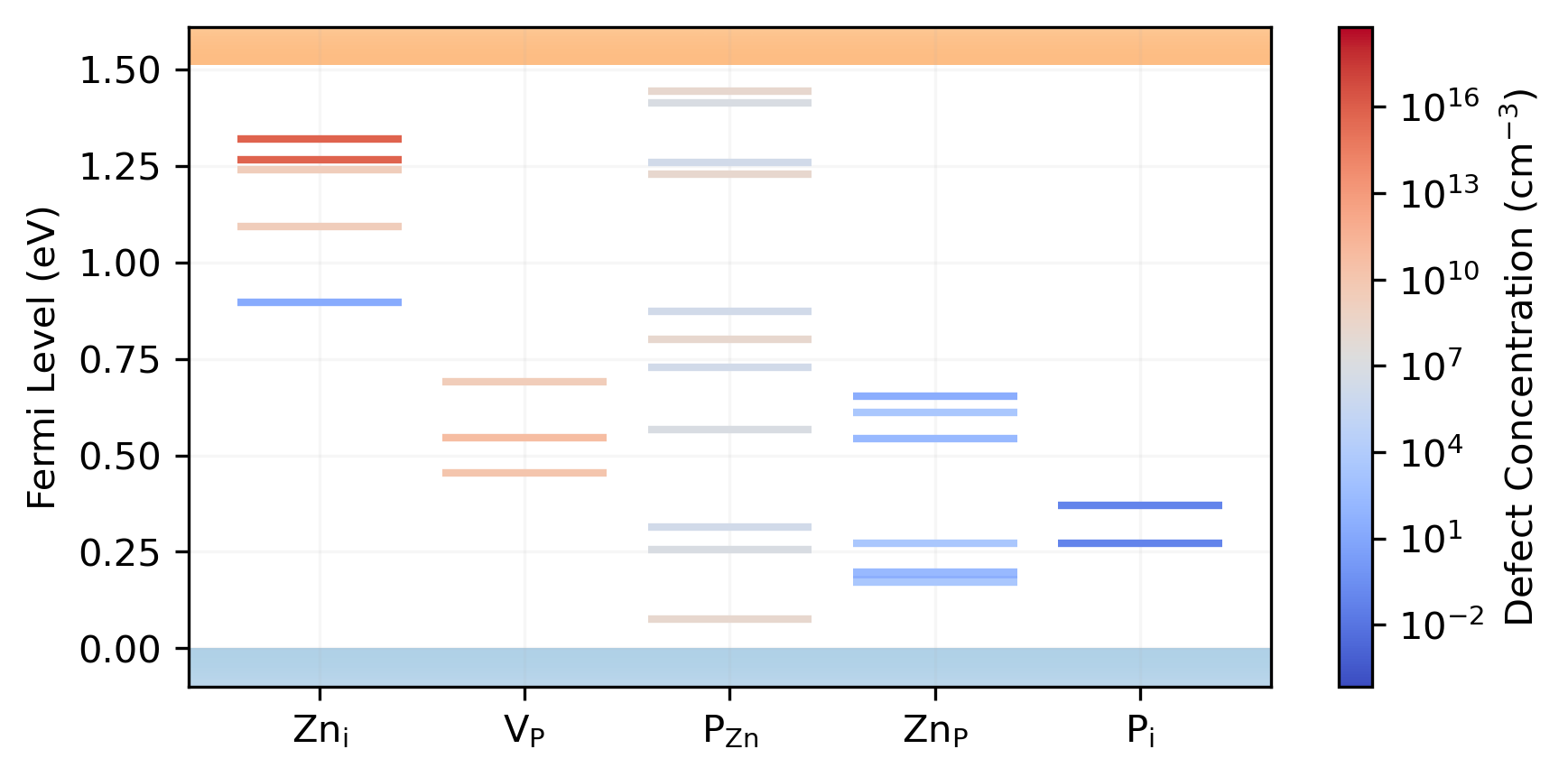}
    \caption{\ Thermodynamic charge‐transition levels of intrinsic point defects in zinc phosphide. The color scale indicates the calculated concentration of each defect species according to the frozen defect approximation (P-rich, $T_\mathrm{anneal}=650\,\mathrm{K}$, $T_\mathrm{quench}=300\,\mathrm{K}$). Only transition levels located within the band gap are shown; zinc vacancies are therefore not included, as all of their computed transition levels lie within the valence band.}
    \label{fig:trans_and_conc}
\end{figure}

In Zn$_3$P$_2$, the most experimentally relevant transitions are expected from the most abundant defects, namely zinc vacancies and zinc interstitials, see Fig.~\ref{fig:trans_and_conc} for the latter. Zinc interstitials show transition levels at about $1.3\,\mathrm{eV}$ above the valence band maximum. This could potentially explain absorption features observed in previous experiments that were historically interpreted as a reduced optical band gap. When defect concentrations are taken into account, transition levels in the range of $1.2$--$1.4\,\mathrm{eV}$ above the valence band maximum are predicted to dominate the defect-related optical response. In contrast, $\mathrm{Zn}$ vacancies appear to exhibit resonant transition levels within the valence band. However, analysis of their orbital character and spatial localization suggests that these electronic states are characteristic of perturbed host states. These hydrogen-like states possess largely extended wavefunctions that span across multiple unit cells, rendering standard supercell correction schemes insufficient. Using a simple hydrogenic effective mass model and given the light effective hole masses reported in Table~\ref{tbl:Zn3P2_pristine_eff_masses}, we estimate that the associated transition levels lie a few meV \emph{above}, rather than below, the VBM, consistent with shallow acceptors with binding energies of a few meV.

A detailed tabulation of various computed thermodynamic transition levels computed in this work is given in Section 5 of the SI.

\subsection{Zinc Frenkel defect}

The dominating presence of oppositely charged zinc interstitials and zinc vacancies suggests a strong Coulomb-driven interaction between these intrinsic point defects. In thermodynamic equilibrium, such oppositely charged defect species are expected to form a bound defect complex, commonly referred to as Frenkel defect. Examining the stability and electronic behavior of these vacancy-interstitial complexes is therefore important for understanding defect and carrier concentrations. 
Frenkel pairs are generally detrimental in semiconductors because the mutual passivation of the vacancy and interstitial charges reduces the density of free carriers, a mechanism well documented in Si and III--V materials.

\begin{figure}
\centering
    \includegraphics[height=3.8cm, trim={0.2cm 0 0.1cm 0}, clip]{
    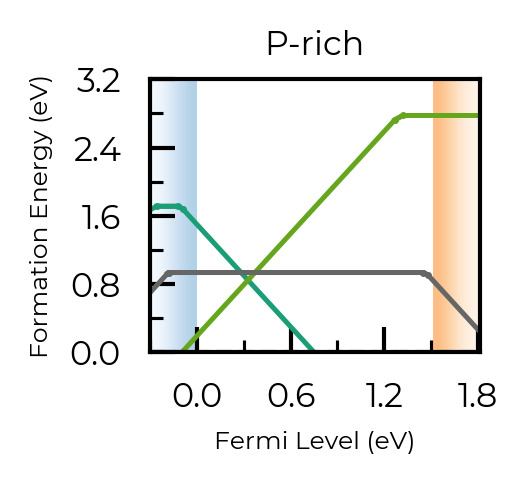}
    \includegraphics[height=3.8cm, trim={0.2cm 0 0.1cm 0}, clip]{
    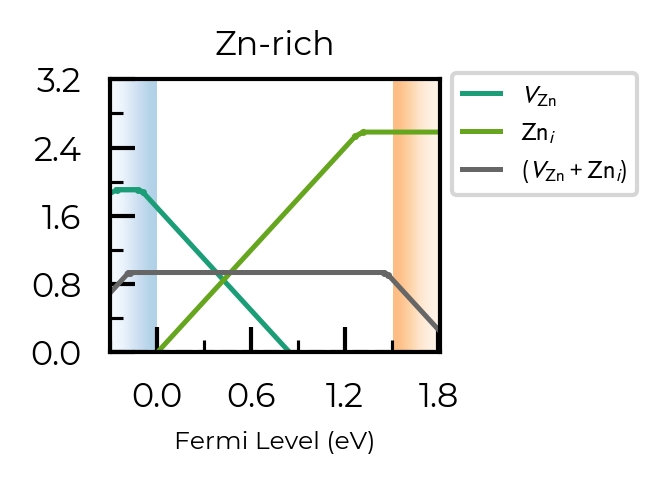}
    \caption{\ Detailed comparison of zinc Frenkel defect components. The transition levels of the isolated zinc vacancy ($V_\mathrm{Zn}$) and zinc interstitial ($\mathrm{Zn}_i$) are shown alongside the $(V_\mathrm{Zn}+\mathrm{Zn}_i)$ complex.}
    \label{fig:intrinsic_pd_transition_energy_diag_frenkel}
\end{figure}

The formation energy of the specifically constructed Frenkel defect complex considered in this work is shown in Fig.~\ref{fig:intrinsic_pd_transition_energy_diag_frenkel}. The complex was generated by placing the lowest-energy configuration found for a zinc interstitial in close proximity to the lowest-energy configuration found for a zinc vacancy, followed by full structural relaxation of all atomic positions. The resulting formation energies are reported under both P-rich and Zn-rich chemical potential limits.

The binding energy of the Frenkel defect is obtained by comparing the formation energy of the bound vacancy-interstitial pair to the sum of the formation energies of the isolated constituents, $\mathrm{V}_\mathrm{Zn}$ and Zn$_i$, at the same chemical potentials and Fermi level. In the present case, the binding energy is effectively independent of the Fermi level within the relevant range, since the Fermi level is restricted between the VBM and approximately $0.85\,\mathrm{eV}$ above it, and neither the isolated defects nor the Frenkel complex undergo charge-state transitions in this interval. Applying eq.~\eqref{equ:binding_energy_complex} yields a positive binding energy of $0.77\,\mathrm{eV}$, indicating a thermodynamically stable, strongly bound defect complex and a significant attractive interaction between the vacancy and interstitial.

With the calculated formation energy, Zn Frenkel defects are predicted to reach concentrations of approximately $10^{16}\,\mathrm{cm}^{-3}$, similar in abundance to zinc vacancies (V$_\mathrm{Zn}$). Frenkel pairs were excluded from the equilibrium Fermi level and carrier concentration computations due to the limitations of the dilute limit approximation. Assuming no interactions between defects, the addition of neutral defect complexes will have no impact on the charge-neutrality condition. Instead, the steady-state concentration of these defects is likely governed by the balance of creation, migration, and recombination, expressed in Kröger-Vink notation as
\begin{equation}
    \mathrm{Zn}_\mathrm{Zn}^\times \rightleftharpoons \mathrm{V}_\mathrm{Zn}'' + \mathrm{Zn}_i^{\bullet\bullet}\,.
    \label{eq:Kroeger-Vink}
\end{equation}
The precise kinetics of these processes depend on diffusion pathways and potential barriers, which remain beyond the scope of this work.

The found Frenkel defects add a new transition energy in the band gap which is slightly closer to the conduction band than that of the $+2$/$0$ charge transition of the isolated zinc interstitial. This could be explained by a stabilizing effect of the interstitial's charge state due to the nearby located negatively charged zinc vacancy. We note, however, that for defect complexes we did not conduct a full search for lower-energy configurations of various charge states. 

\subsection{Comparison to previous studies}

The defect chemistry of Zn$_3$P$_2$ has been the subject of several evolving computational studies. Early assessments by Demers \textit{et al.}~\cite{demers_intrinsic_2012} utilized semilocal functionals, which necessitated ad hoc band gap corrections that introduced significant uncertainty into the defect energetics. Subsequently, Yin \textit{et al.}~\cite{yin_electronic_2013} improved upon this using hybrid functionals, though their results differed from later studies, probably due to a lack of spin polarization and finite-size corrections~\cite{yuan_first-principles_2023}.

The most recent comprehensive study by Yuan \textit{et al.}~\cite{yuan_first-principles_2023} addressed many of these functional-level deficiencies, predicting that zinc vacancies (V$_\mathrm{Zn}$) act as a shallow acceptor and that phosphorus interstitials P$_i$ give rise to deep levels. Our results align with some of the trends established in Ref.~\cite{yuan_first-principles_2023}, yet several critical deviations emerge due to differences in computational methodology and our more exhaustive search of the configuration space and charge state stability.

\begin{figure}
\centering
    \includegraphics[width=1.0\linewidth]{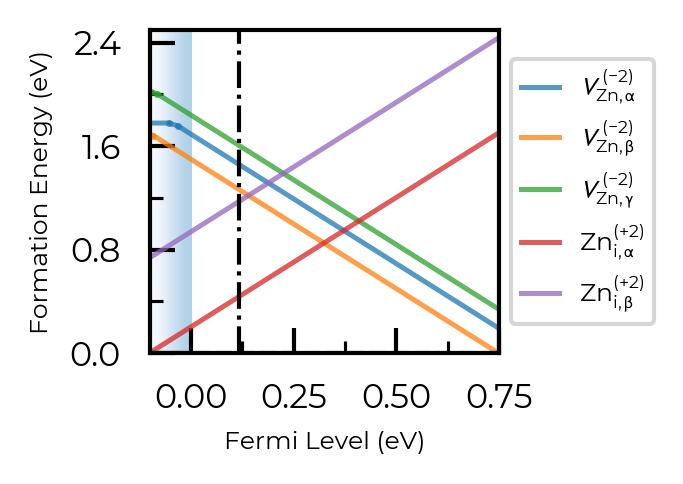}
    \caption{\ Closeup of the transition energy diagram around the equilibrium Fermi level at $0.12\,\mathrm{eV}$. Shown are stable and metastable configurations of zinc vacancies and zinc interstitials, the two dominating types of defects.}
    \label{fig:intrinsic_pd_transition_energy_diag_closeup}
\end{figure}

\begin{figure}
\centering
    \includegraphics[width=0.48\linewidth]{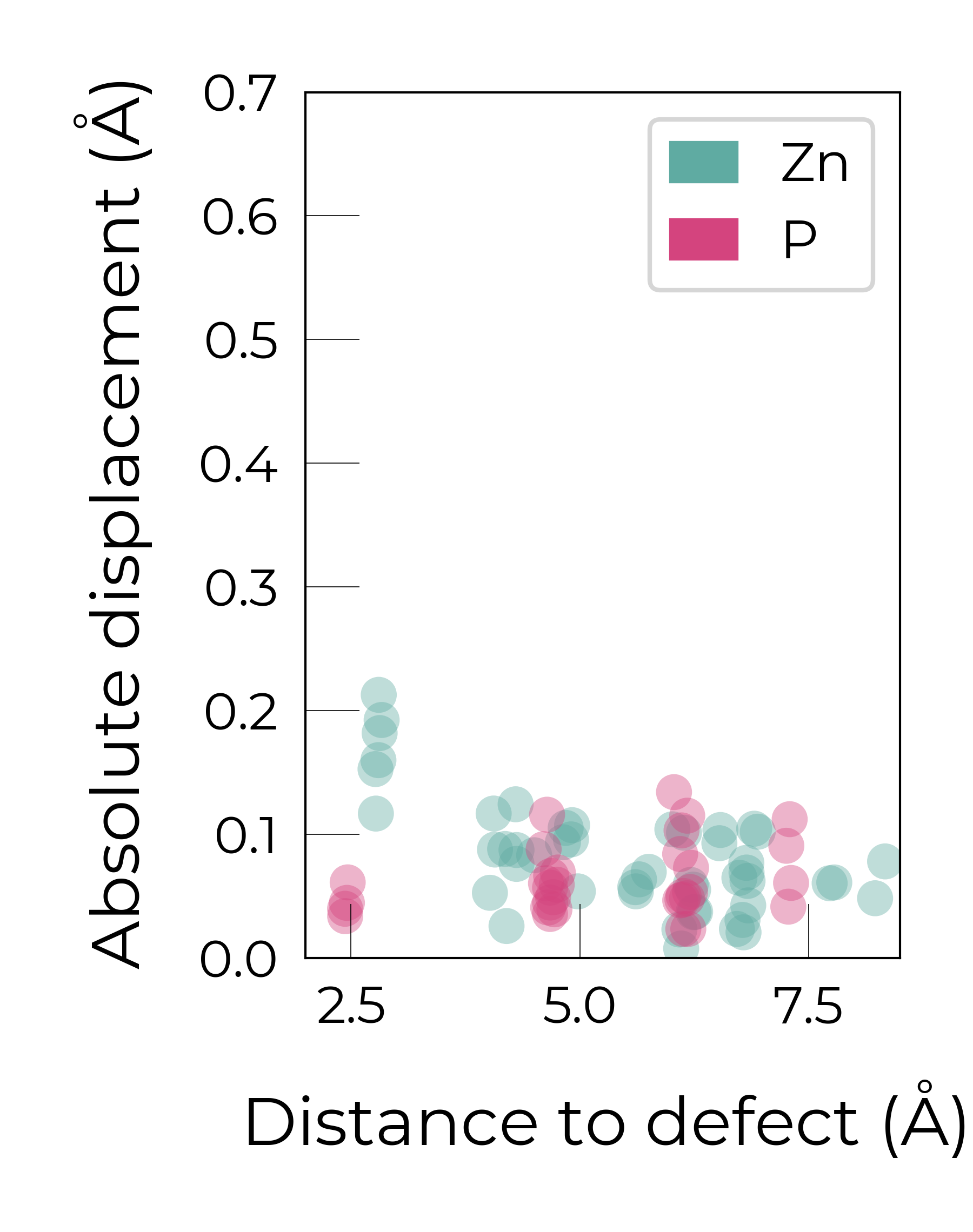}
    \includegraphics[width=0.48\linewidth]{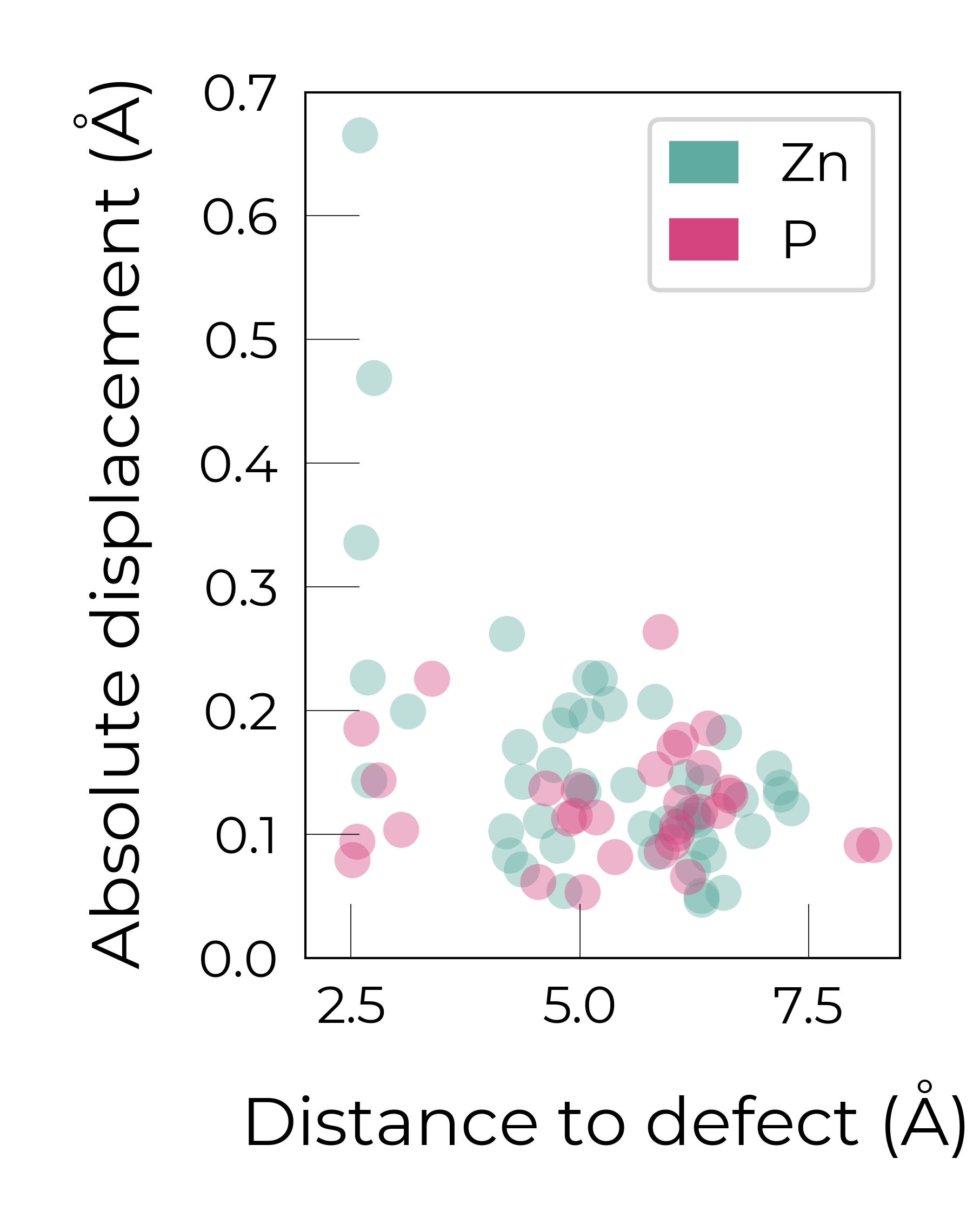}
    \caption{\ Absolute atomic displacements as a function of distance from the defect center for the two lowest-energy zinc interstitial configurations Zn$_{i,\alpha}$ (left) and Zn$_{i,\beta}$ (right).}
    \label{fig:Zn_i_displacement_plots}
\end{figure}

The primary distinction in the present work lies in the identification of ground-state configurations. While Yuan \textit{et al.}\ performed robust local optimizations, our efforts to identify several local minima and enforce explicit symmetry-breaking revealed significantly lower-energy configurations for both the zinc interstitial (Zn$_i$) and the zinc vacancy (V$_\mathrm{Zn}$), see Fig.~\ref{fig:intrinsic_pd_transition_energy_diag_closeup}. Specifically, we find that the lowest-energy Zn$_i$ configuration occupies a site at the center of a tetrahedron formed by P-atoms (rather than Zn-atoms as previously assumed), accompanied by a significantly smaller distortion of the surrounding lattice, with atomic displacements decaying rapidly within 5\,\AA\ of the defect center as shown in Fig.~\ref{fig:Zn_i_displacement_plots}.

Furthermore, Yuan \textit{et al.}\ chose to tune the amount of mixing in the HSE functional family such that the experimental band gap is reproduced, as has been suggested for computations of point defects~\cite{fried_electronic_2026}. We chose to stick to the standard value of $\alpha =  0.25$. In fact, while tuning $\alpha$ can improve band gap agreement, it does so at the cost of total-energy accuracy. Formation energies and transition levels are more sensitive to errors in the total energies of the defect supercells than to the precise position of the band edges, and standard HSE06 has been shown to yield reliable defect formation energies across a broad range of semiconductors even when the band gap is moderately underestimated~\cite{freysoldt_first-principles_2014}. We therefore chose to remain agnostic with regard to the band gap, while relying on well-validated methods for energetics. A post-calculation scissoring shift to E$_\mathrm{gap}= 1.51\,\mathrm{eV}$ was applied to the final transition level diagrams to facilitate comparison with literature.

\begin{figure}
\centering
    \includegraphics[width=1.0\linewidth]{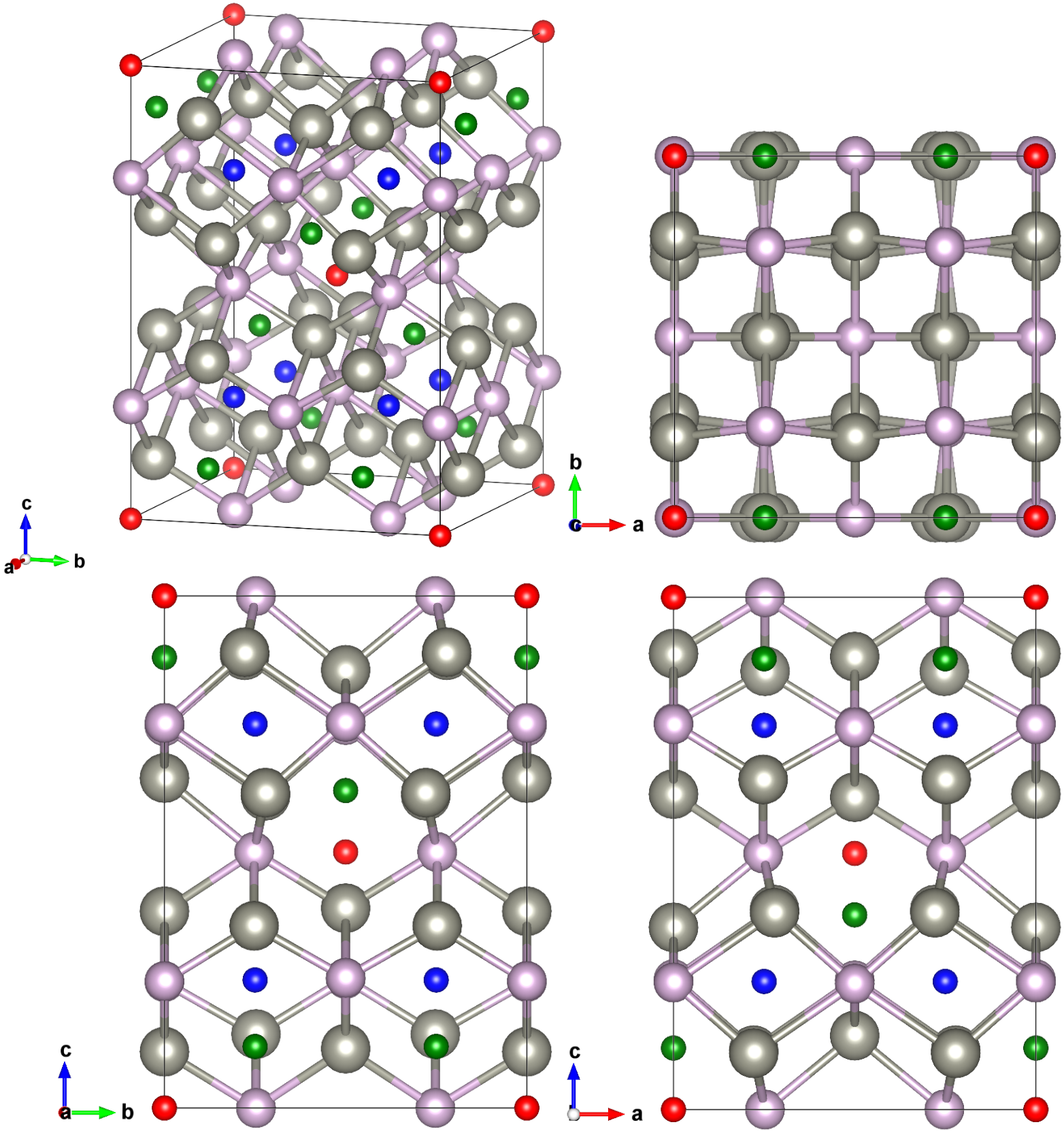}
    \caption{\ Different views of the conventional unit cell of Zn$_3$P$_2$, illustrating the approximate positions of selected Zn$_i$ configurations. Colored spheres (blue, red, and green) were manually inserted into the pristine unit cell without subsequent geometry optimization and are shown for visualization purposes only. Multiple occurrences of the same-colored defect indicate symmetry-equivalent positions in the pristine crystal structure rather than distinct defect sites. See the main text for more information.}
    \label{fig:Zn_i_in_conventional_cell}
\end{figure}

Figure~\ref{fig:Zn_i_in_conventional_cell} illustrates several considered Zn$_i$. The configuration shown in red corresponds to the previously proposed lowest-energy site~\cite{yuan_first-principles_2023}, where the interstitial Zn-atom occupies the center of a tetrahedron formed by Zn-atoms and an octahedron formed by P-atoms. The green configuration represents the lowest-energy Zn$_i$ site identified in this work; here, the interstitial resides at the center of a tetrahedron formed by P-atoms and an octahedron formed by Zn-atoms. Finally, the configuration shown in blue corresponds to the second-lowest-energy Zn$_i$ site found in this study, which lies energetically close to the red configuration and is characterized by an octahedral coordination formed by both P- and Zn-atoms. Viewed in terms of the (001) atomic layers of the pristine conventional unit cell, the lowest-energy interstitial configuration (green) occupies a site corresponding to a naturally occurring zinc ``vacancy'' (when comparing to a fluorite structure) in the Zn sublattice. In contrast, the red and blue configurations are located within different P-layers, previously believed to be more suited as hosts for Zn interstitials.

We find the observed energetic ordering to be primarily driven by electrostatic interactions. Specifically, the Ewald energy of the lowest-energy configuration (green) is more than $50\,\mathrm{meV/atom}$ lower than that of the second-lowest-energy configuration (blue). This difference suggests that an increased ratio of Zn-P nearest-neighbor interactions facilitates a more energetically favorable local environment.

Furthermore, our results indicate that phosphorus interstitials P$_i$ are considerably higher in formation energy than previously reported. This shift suggests that the P$_i$ plays a much less prominent role in the equilibrium defect chemistry of Zn$_3$P$_2$ than suggested by either Demers \textit{et al.}\ or Yuan \textit{et al.}, and most likely do not significantly increase detrimental recombination due to deep levels.

By exploring a wider range of charge states than previous studies, we also find low-energy configurations for P$_\mathrm{Zn}$ $-3$ and $-4$ charge states, which add previously unreported transition levels in the band gap. Additionally, we have extended the scope of inquiry to include defect-defect interactions. Our findings show a strong attractive interaction between zinc interstitials and vacancies, suggesting that defect clustering may be a significant factor.

During the preparation of this manuscript, a study by Zhang \textit{et al.}~\cite{zhang_theoretical_2026} was published, which also addresses intrinsic and extrinsic point defects in Zn$_3$P$_2$. While the scopes of these investigations overlap, significant differences exist in the computational treatment and the reporting of methodological parameters. The present work provides a more robust supercell model, exhaustive account of functional selection, finite-size correction, and the derivation of thermodynamic properties, including equilibrium Fermi levels and defect concentrations.

\section{Summary and Conclusion}

This work examines the ground-state energetics and thermodynamics of intrinsic point defects in Zn$_3$P$_2$.

Our defect thermodynamics analysis shows that zinc vacancies and zinc interstitials are the dominant intrinsic point defects, with equilibrium concentrations several orders of magnitude higher than those of all other defects considered. In contrast, phosphorus interstitials, which were previously proposed as abundant, are predicted to occur only in negligible concentrations due to their high formation energies, even under phosphorus-rich growth conditions, which rather facilitates the formation of P$_\mathrm{Zn}$ antisites.
%
The intrinsic $p$-type conductivity of nominally undoped Zn$_3$P$_2$ emerges naturally from the prevalence of zinc vacancies. These defects are thermodynamically stable in negative charge states and act as electron acceptors, thereby depleting free electrons and shifting the Fermi level toward the VBM.
%
We further find that zinc vacancies and zinc interstitials, the two most abundant intrinsic defects, exhibit a positive binding energy and therefore tend to form neutral Frenkel pairs. The formation of such defect pairs is expected to partially compensate the effect of isolated zinc vacancies and thus moderates the intrinsic $p$-type conductivity.

From a device-design perspective, controlling carrier concentrations is essential. While we find no intrinsic mechanism that yields substantial free-electron concentrations, hole densities can be significantly enhanced by favoring zinc-vacancy formation. This can be achieved through growth under phosphorus-rich conditions (e.g., elevated phosphorus partial pressures in molecular beam epitaxy) combined with high growth temperatures, or through post-growth annealing at elevated temperatures.

While our results provide a good model for the experimentally observed $p$-type character, we note that this work considers only intrinsic point defects. In real samples, extrinsic impurities or extended defects such as grain boundaries can also affect the electronic properties. Nevertheless, we found little evidence in the literature for such significant competing effect and believe our analysis captures most of the fundamental defect thermodynamics of the pristine material.

The dominance of zinc vacancies among intrinsic point defects explains the persistent difficulty of achieving reliable $n$-type conductivity and indicates that such behavior is not primarily limited by synthesis quality, but rather by thermodynamic constraints. Consequently, any extrinsic donors compete with compensation unless intrinsic defect populations can be significantly altered.
%
Rather than pursuing bulk $n$-type doping, future photovoltaic device strategies may benefit more from further enhancing hole concentrations and from approaches such as interface engineering and heterostructure design to control carrier transport and recombination.

\section*{Conflicts of interest}

There are no conflicts of interest to declare.

\section*{Acknowledgments}

N.K. and S.B. acknowledge funding from the European Innovation Council and SMEs Executive Agency (EISMEA) under grant agreement No 101046297 (Pathfinder Project SOLARUP). The authors thank the colleagues of the SOLARUP consortium for fruitful scientific discussions and for their critical review of the manuscript. 
%
The authors gratefully acknowledge the computing time provided to them on the high-performance computers Noctua 1 \& 2 at the NHR Center PC2. These are funded by the Federal Ministry of Education and Research and the state governments participating on the basis of the resolutions of the GWK for the national high-performance computing at universities (www.nhr-verein.de/unsere-partner).

The authors gratefully thank Yuan \textit{et al.}~\cite{yuan_first-principles_2023} for kindly providing the structural models used in their point-defect calculations.

Crystal structures were visualized with \texttt{VESTA}~\cite{momma_vesta_2008}. We made use of the \texttt{sumo} package~\cite{ganose_sumo_2018} for post-processing and analysis of the DFT calculations. Various 2D-plots in this work were drawn using \texttt{matplotib}~\cite{hunter_matplotlib_2007}.







\bibliography{PAPER_Zn3P2_IPD,rsc} 
\bibliographystyle{rsc} 
\end{document}